\documentclass[english,aps,prb,showkeys,reprint]{revtex4-2}

\usepackage{amsmath}
\usepackage{amssymb}
\usepackage{amsfonts}
\usepackage{xfrac}
\usepackage{accents} 
\usepackage{graphicx}
\DeclareGraphicsExtensions{.pdf,.png,.jpg}
\usepackage[dvipsnames]{xcolor}
\usepackage{verbatim} 
\usepackage{multirow}

\newcommand{\dd}{ \mathrm{d} }
\newcommand{\be}{\begin{equation}}
\newcommand{\ee}{\end{equation}}

\begin{document}

\title{A Rate Equation for the Transfer of Interstitials \\ across Interfaces between Equilibrated Crystals}

\author{Jörg Weissmüller}

\affiliation{Institute of Materials Physics and Technology, Hamburg University of Technology, Hamburg, Germany and Institute of Hydrogen Technology, Helmholtz-Zentrum Hereon, Geesthacht, Germany}

\date{\today}

\begin{abstract}
This work inspects the thermally activated transfer of solute particles across the interface between two interstitial solid solution phases that equilibrate internally by fast diffusion on conserved arrays of sites. When each phase is considered as an ergodic ensemble of particles, statistical mechanics predicts the occupancy of the transition states at equilibrium to depend on the barrier energy and on the chemical potentials and vacancy fractions in each of the phases. A rate law for the non-equilibrium interfacial transfer, based on a constant transition probability between activated states, naturally satisfies the principle of detailed balance. Contrary to Butler-Volmer-type laws, values of the particle chemical potentials enter explicitly rather than through their difference. This, along with the dependency on the vacancy fractions, implies here an exchange flux density that depends explicitly on the compositions at equilibrium. The results can explain experimental observations of a drastic slow-down in the charging of metal hydrides near phase transformations or miscibility-gap critical points.
\end{abstract}

\keywords{kinetic rate theory, metal hydrides, detailed balance, Butler-Volmer law, statistical mechanics}

\maketitle

\textit{Introduction}
--
This work is motivated by observations on the characteristic time, $\mathcal T$, for composition changes in interstitial metal hydride nanomaterials when they exchange solute with an external reservoir at controlled chemical potential. The observations tie to the larger field of interstitial solid solutions as energy storage materials, comprising also lithium compounds as battery electrodes, where the kinetics of charging/discharging underlies key figures of merit for application. In hydrides of Pd and PdPt, $\mathcal T$ can increase by one to several orders of magnitude when the interval of composition change includes a phase transformation or the vicinity of the critical point of the miscibility gap \cite{ShanShi2017,BapariWeissmüller2025}. The Butler-Volmer (BV) equation \cite{Dickinson2020,Fletcher2023}, which serves as a basis for analyzing the kinetics of electrochemically controlled hydriding \cite{Montella1996,Pyun2005, Montella2005, WuerschumPdH2024}, does not provide an obvious explanation of that observation.

The electrochemically controlled insertion of H in Pd or Pd-based alloys starts out with adsorption by the Volmer reaction \cite{Lasia2019}, $\rm H^+ + e^- \rightleftharpoons H_{ads}$, and proceeds by interfacial transfer into the bulk absorbed state \cite{Jerkiewicz1998}. It has been suggested that overcoming an energy barrier between the adsorbed and absorbed states provides the rate-limiting step for the insertion \cite{BapariWeissmüller2025}. With the charge transfer completed during the Volmer reaction, transfer across the barrier involves no electron transfer, contrary to the redox process that underlies the Butler-Volmer equation \cite{BockrisKhan1993,BardFaulkner2001}. Two further characteristics of hydrogen insertion are also not immediately exposed in that equation. Firstly, explaining the variation of $\mathcal T$ with the hydrogen fraction, $x$, requires that one explicitly acknowledges the presence of a continuum of equilibrium states, of different $x$, that are selected by the electrode potential. Secondly, as a consequence of the misfit strain energy, interstitial solid solutions inherently feature a strong, attractive solute-solute interaction, which affects the driving forces and often results in a miscibility gap \cite{WeissmuellerInterstitial2024}.

Internal barriers in nanoscale metal hydrides are characterized by yet another distinctive feature. The characteristic time for internal equilibration in a nanoscale region of the bulk can be orders of magnitude less than the time, namely $\mathcal T$, for equilibration with the external reservoir.
Take Pd at 300K -- the H bulk diffusion constant $\rm 4 \times 10^{-11} m^2/s$ \cite{FlanaganOates1991} implies internal equilibration of 10 nm particles within $< 1 \mu s$ while $\mathcal T$ can be minutes \cite{ShanShi2017,Dionne2018,BapariWeissmüller2025}.
The thermally activated jumps that govern $\mathcal T$ may then be viewed as rare transfer events between extended ensembles of solute particles that undergo fast, ergodic sampling of their configuration space. Interstitial sites separated by the barrier will here accommodate solute at well-defined chemical potentials, while their occupancies fluctuate. Due to the solute-solute interaction, the ground-state energies of particles in the sites will also fluctuate. This raises the question if the scenarios at hand are correctly represented by rate laws -- such as the BV equation -- based on jump probabilities between sites with fixed a priori occupancy and fixed ground state energy.

Here, we derive a kinetic rate law for the thermally activated transfer of charge-neutral particles between two ergodic interstitial solution phases. Statistical mechanics implies the transition state occupancy at equilibrium for each phase. The transfer between the phases arises from transitions between their respective activated states, with differences in the chemical potentials providing the driving force for a net particle flux. The ensuing rate law naturally satisfies the principle of detailed balance, and it explains qualitatively the experimentally observed slowdown near the critical point.

\textit{Fundamentals and constitutive assumptions}
--
We consider a composite system, $\sf S$, comprising two interstitial solid solution phases, $\sf A$ and $\sf B$. The phases mutually exchange particles of solute by thermally activated transfer across an interfacial energy barrier that accommodates a transition state, $\sf T$. Transfer across the barrier is a rare event, the rate-limiting step for the equilibration. Bulk diffusion is considered so fast that each phase individually reaches internal equilibrium quasi instantaneously after insertion or extraction of solute. We may then assume ergodic sampling of the configuration space in each phase.
In-between particle exchange events, $\sf S$ may also be considered as an ergodic system, equilibrated subject to the constraint of constant amounts of particles and sites, separately in each of the two phases.

We take $\sf S$, $\sf A$ and $\sf B$ as canonical systems, at constant and uniform temperature, $T$. Ignoring capillarity, we take all thermodynamic potentials as homogeneous first order functions.

Interstitial solid solutions are modeled as conserved networks of sites that can be empty or occupied by a single solute particle. Phases are regions of the network; each phase contains $\mathcal{N}_{\rm S}$ particles on $\mathcal{N}_{\rm M}$ sites. Solute fractions are defined as $x = \mathcal{N}_{\rm S} / \mathcal{N}_{\rm M}$.

\begin{figure} [!t]
	\centering
	\includegraphics[width=0.6\columnwidth]{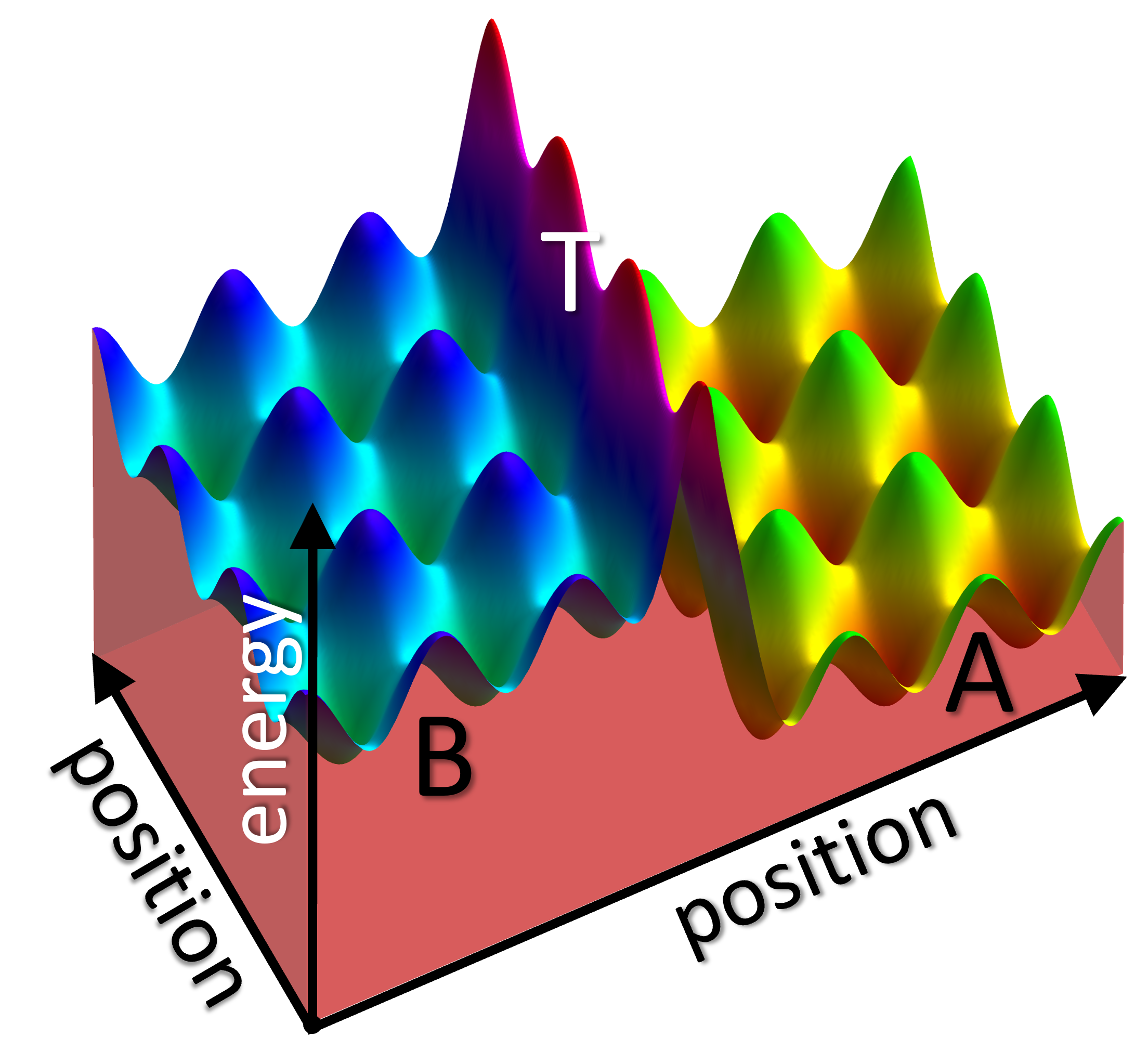}
	\caption{Scheme of energy landscape, variation of single-particle potential energy with position. Phases $\sf A$ and $\sf B$ exhibit periodic energy variation, with minima defining sites available to particles. Phases are separated by an energy barrier, and transition states $\sf T$ represent local saddle points between neighboring sites on opposite sides of the barrier.}
	\label{fig:energy_landscape}
\end{figure}

The transition state is defined in a single-particle potential energy landscape (Fig \ref{fig:energy_landscape}), where $\sf T$ forms a saddle point at the interface, in-between two neighboring sites, the transfer sites, one on each side of the interface. Transfer events require a "transfer configuration", $\sf C$, that has a particle in a transition state, at energy $\epsilon^{\sf T}$, in the source site and an empty target site.

In a description by phenomenological thermodynamics \cite{CallenThermodynamicsBook}, and restricting attention to scenarios with negligible work of mechanical deformation, each phase may be characterized by an inner energy function $\mathcal{U}(\mathcal{S}, \mathcal{N}_{\rm S}, \mathcal{N}_{\rm M})$ with the fundamental equation
\be
\label{eq:Inner_energy_FE}
\dd \mathcal{U}
=
T \dd \mathcal{S}  +  \mu_{\rm S}  \dd  \mathcal{N}_{\rm S}  +  \mu_{\rm M} \dd  \mathcal{N}_{\rm M} \,.
\ee

Here, $\mathcal{S}$ denotes the entropy, and $\mu_{\rm S}$, $\mu_{\rm M}$ define the chemical potentials of solute particles and matrix sites, respectively.
The condition for two-phase equilibrium, with respect to particle exchange between phases $\sf A$ and $\sf B$, is
\be
\label{eq:mu_equilibrium}
\mu_{\rm S}^{\sf A}
=
\mu_{\rm S}^{\sf B}  \,.
\ee
Free energies, $\mathcal{F}$, are defined as
\be
\label{eq:F_definition}
   \mathcal{F} = \mathcal{U} - T \mathcal{S} \,.
\ee
Ergodicity implies that thermodynamic potentials such as $\mathcal{U, S, F}$ can be defined and take on precisely defined values.

In a statistical mechanics description \cite{CallenThermodynamicsBook,Chandler1987}, the entropy of a system is
\be
\label{eq:Entropy_definition}
\mathcal{S} = k_{\rm B} \ln \Omega \,,
\ee
with $\Omega$ the number of microstates and $k_{\rm B}$ Boltzmann's constant. The partition sum,
\be
\label{eq:Partition_function}
Z = \sum_j \exp -\frac{E_j}{k_{\rm B} T}\,,
\ee
relates to the free energy by
\be
\label{eq:Free_energy_Z}
\mathcal{F} = - k_{\rm B} T \ln Z\,.
\ee
For a canonical system with conserved amounts of solute and matrix sites, the probability of a single microstate $j$ with energy $E_j$ is \cite[§16]{CallenThermodynamicsBook}\cite[§31]{LandauLifschitzBd5_1991}
\be
\label{eq:Boltzmanns_equation}
W_j =
\exp -\frac{E_j - \mathcal{F}}{k_{\rm B} T} \,.
\ee
Here, $E_j$ refers to the energy of the entire system in one of its microstates; this energy is distinct from the per-particle energy level $\epsilon^{\sf T}$ of an individual solute atom in a transition state.

\textit{Constrained equilibrium and probability of the transfer configuration}
--
We now compute $W_{{\sf C}_p}$, the probability of finding $\sf S$ in a transfer configuration, ${\sf C}_p$, with an activated particle in source site $p$. Without lack of generality we consider that site located in $\sf A$.
We treat $W_{{\sf C}_p}$ as a property of $\sf S$ \emph{at equilibrium} (again subject to the constraint of fixed $\mathcal N$ in each of the two phases).

We emphasize the conceptual distinction between, on the one hand, the activated local state of a single particle at the transfer site, of (single-particle-) energy $\epsilon^{\sf T}$, and, on the other hand, microstates of the entire composite system $\sf S$ that have ${\sf C}_p$ activated. While we consider only one local configuration of the activated particle, there will generally be many microstates of $\sf S$ that include that local configuration.

With the last-mentioned observation in mind, we compute $W_{{\sf C}_p}$ by summing over the per-microstate probabilities, $W_{{\sf C}_{p,j}}$, of all microstates $j$, with energies $E_j$, within ${\sf C}_p$:
\begin{align}
\nonumber
W_{{\sf C}_p}
&=
\sum_j W_{{\sf C}_{p,j}}
=
\sum_j \exp{ -\frac{E_j - \mathcal{F}}{k_{\rm B} T}} \\
&=
  Z_{{\sf C}_p} \exp\frac{\mathcal{F}}{k_{\rm B} T} \,,
\label{eq:W_total_Sum}
\end{align}
where, in view of Eq \ref{eq:Partition_function}, $ Z_{{\sf C}_p}$ is the partition sum of the system when constrained to have ${\sf C}_p$ activated. Considering Eq \ref{eq:Free_energy_Z}, we see that $ Z_{{\sf C}_p}$ implies $ \mathcal{F}_{{\sf C}_p}$, the free energy of the constrained system, in terms of which Eq \ref{eq:W_total_Sum} reads
\be
\label{eq:W_total_Sum_F}
W_{{\sf C}_p}
=
\exp{-\frac{ \mathcal{F}_{{\sf C}_p} - \mathcal{F}}{k_{\rm B} T}}\,.
\ee
As compared to Eq \ref{eq:Boltzmanns_equation}, the exponent in Eq \ref{eq:W_total_Sum_F} no longer contains the energy of a specific microstate. Instead, it is governed entirely by thermodynamic averages in the form of free energies.

\begin{figure} [!t]
	\centering
	\includegraphics[width=0.90 \columnwidth]{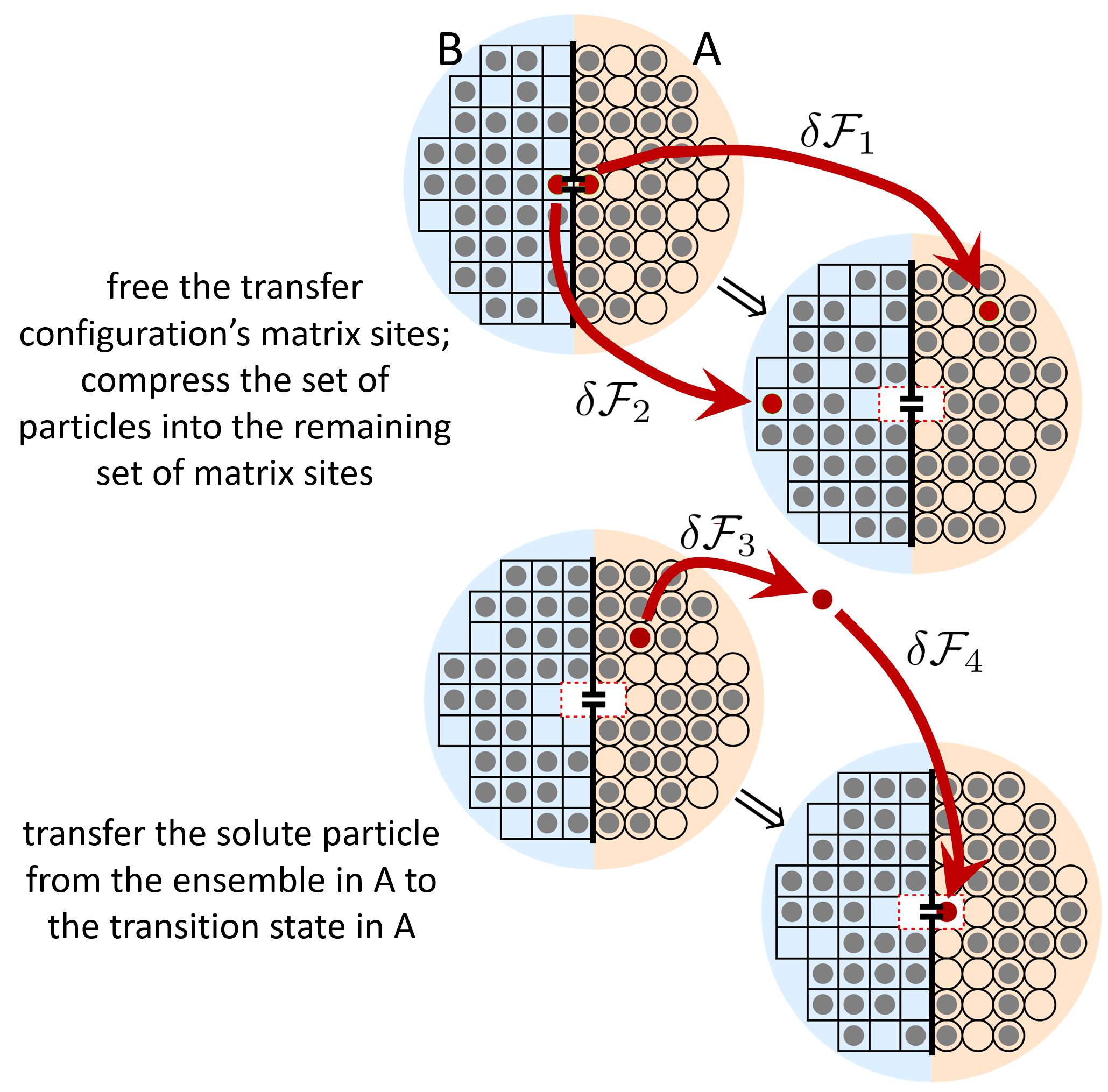}
	\caption{Virtual steps in computing the transfer configuration free energy. Sites in phases $\sf A$ (circles, orange shade ) and $\sf B$ (squares, blue shade) are occupied by solute particles (gray discs). Top: freeing the transfer sites by compressing the respective sets of particles to the smaller sets of remaining sites. Bottom: removing a particle (red disk) from the bulk ensemble in $\sf A$ and inserting it into the source site of the transfer configuration ("=" and white rectangle). Red arrows correspond to the four $\delta \mathcal{F}$ in the main text.}
	\label{fig:transfer}
\end{figure}

The free energy $\mathcal{F}_{{\sf C}_p}$ differs from $\mathcal{F}$ since the occupancies of the two sites in the transfer configuration are fixed -- occupied source site, vacant target site -- and since the particle in the source site is in an activated state. For comprehensibility we separate the computation of the free energy change, $\delta \mathcal{F} = \mathcal{F}_{{\sf C}_p} - \mathcal{F}$, into discrete steps (see Fig \ref{fig:transfer}).

Step 1 makes sure that the transfer source site is vacant, starting out with removing that site from the ergodic ensemble in phase $\sf A$. As a consequence, the $\mathcal{N}^{\sf A}_{\rm S}$ particles are compressed into the $\mathcal{N}^{\sf A}_{\rm M} - 1$ remaining matrix sites. Next, the source site is reinserted at its original position, while keeping the solute constrained to the smaller ensemble of sites. Since the net result of the two substeps is simply the redistribution of the solute on one fewer sites, the free energy change reflects the ensuing variation in the configurational entropy of mixing, $S^{\sf A}_{\rm con}$. The free energy change then $\delta \mathcal{F}_1 = T \sigma^{\sf A}_{\rm M}$ with $\sigma_{\rm M} = \dd S_{\rm con} / \dd \mathcal{N}_{\rm M} |_{T, \, \mathcal{N}_{\rm S}}$.

Step 2 is the analogous process on the target side, with a free energy change $\delta \mathcal{F}_2 = T \sigma^{\sf B}_{\rm M}$.

Step 3 removes a solute particle from the ensemble in $\sf A$ and transfers it into an (arbitrary) reference state where $\mu_{\rm S} = \mu^0_{\rm S}$; this brings $\delta \mathcal{F}_3 = \mu^0_{\rm S} - \mu^{\sf A}_{\rm S}$.

Lastly, step 4 transfers the solute particle from the reference state into the specific transition state at the source site that was vacated in step 1. It brings $\delta \mathcal{F}_4 = \epsilon^{\sf T} - \mu^0_{\rm S}$.

Summing up the free energy changes associated with all steps, the net free energy change is obtained as
\be
\label{eq:sumup_Fp}
\mathcal{F}_{{\sf C}_p} - \mathcal{F} = T \sigma^{\sf A}_{\rm M} + T \sigma^{\sf B}_{\rm M} - \mu^{\sf A}_{\rm S} + \epsilon^{\sf T} \,.
\ee
Inserting Eq \ref{eq:sumup_Fp} into Eq \ref{eq:W_total_Sum_F} yields the probability of finding ${\sf C}_p$ activated,
\be
\label{eq:W_total_closed_explicit_first}
W_{{\sf C}_p}
=
\exp - \frac{\sigma^{\sf A}_{\rm M} + \sigma^{\sf B}_{\rm M}} {k_{\rm B}}
\exp - \frac{\epsilon^{\sf T} - \mu^{\sf A}_{\rm S}}{k_{\rm B} T}  \,.
\ee
The probability of finding a vacancy at a specific site from the array of identical ones in the uniform interstitial solution (which is the configuration of steps 1 and 2) is $1 - x$. Thus, one can rewrite Eq \ref{eq:W_total_closed_explicit_first} as \cite{Note1}
\be
\label{eq:W_total_closed_explicit}
W_{{\sf C}_p}
=
(1 - x^{\sf A}) (1 - x^{\sf B})
\exp - \frac{\epsilon^{\sf T} - \mu^{\sf A}_{\rm S}}{k_{\rm B} T}  \,.
\ee
Importantly, $ W_{{\sf C}_p}$ at two-phase equilibrium (where the $\mu_{\rm S}$ agree in the two phases, Eq \ref{eq:mu_equilibrium}) emerges as independent of the identity of the source phase, $\sf A$ or $\sf B$.

\begin{figure} [!t]
	\centering
	\includegraphics[width= 0.7 \columnwidth]{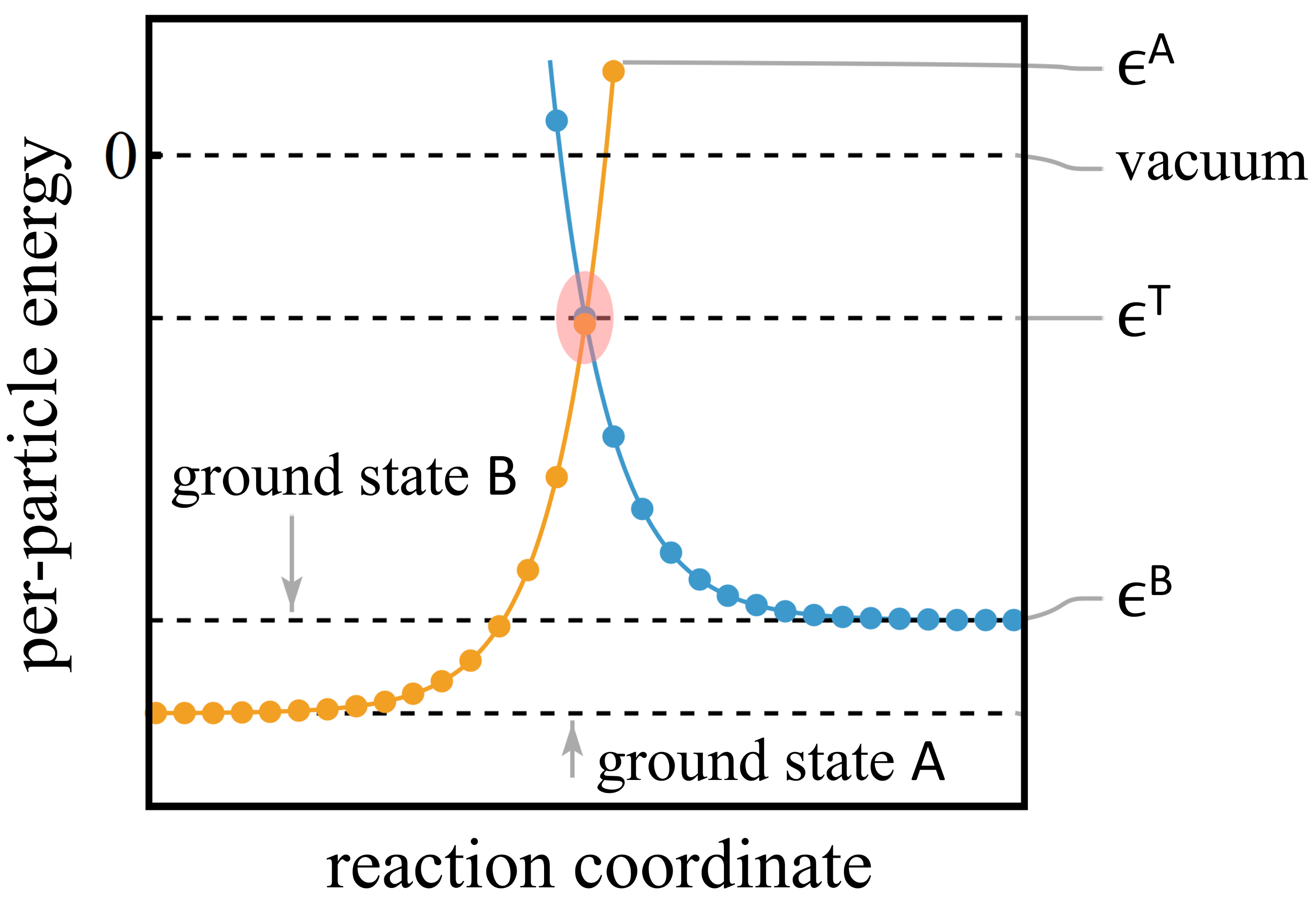}
	\caption{Scheme of per-particle energy $\epsilon$ versus reaction coordinate for transitions between phases $\sf A$ and $\sf B$ across a barrier at transition state $\sf T$.}
	\label{fig:3change-l}
\end{figure}

\textit{Transition rate and flux}
--
By analogy to the analysis of electron transfer reaction rates for redox processes \cite{BardFaulkner2001}, each pair of transfer sites may be associated with two separate transition states, one for each phase. As illustrated  in Fig \ref{fig:3change-l}, we may think of "reaction coordinates" for particles, separately in each of the two phases, and we may envisage these coordinates to overlap at the per-particle energy level $\epsilon^{\sf T}$ \cite{BardFaulkner2001,VanSanten2009}. Taking identical per-particle energy levels for the transition states results in a radiationless transition during the transfer of a particle \cite{BockrisKhan1993}.

A transfer event is initiated when a particle changes over between the transition states of the two adjacent phases. The particle may then relax to the ground state and be collectivized in its new phase $\sf B$, concluding the transfer event. We take the flux density (atoms transferred per area per time) of solute from phase ${\sf A}$ into phase ${\sf B}$ as
\be
\label{eq:Flux_alpha_beta}
J^{\sf AB} = \nu_0 \rho_{\sf T} W_{\sf C}^ {{\sf A}}\,.
\ee
Here, $\nu_0$ is an exchange frequency, $\rho_{\sf T}$ the density (per area of interface) of transfer site pairs, and $W_{\sf C}^{{\sf A}}$ is the probability of the transfer configuration with the source site in ${\sf A}$. Note that our derivation of the $W_{\sf C}$ ensures that the transition state of the receiving phase in $\sf C$ is always empty (available).

The value of the exchange frequency $\nu_0$ is outside the scope of our considerations. We assume this frequency to be sensibly constant throughout the regions of parameter space examined here.

Next, we designate by $J^{\sf B}$ the net flux density into $\sf B$. This quantity is the difference between the in- and out-fluxes; with Eq \ref{eq:Flux_alpha_beta} it results as
\be
\label{eq:Flux_net}
  J^{\sf B}
  =
  J^{\sf AB} - J^{\sf BA}
  =
  \nu_0 \rho_{\sf T}
\left(
W_{\sf C}^ {\sf A} - W_{\sf C}^ {\sf B}
\right)\,.
\ee
Combining Eqs \ref{eq:W_total_closed_explicit} and \ref{eq:Flux_net}, we obtain the rate equation in explicit form as
\be
\begin{split}
  \label{eq:Flux_net_fundamental}
  J^{\sf B}
  &=
\nu_0 \rho_{\sf T}
(1 - x^{\sf A}) (1 - x^{\sf B})
  \exp - \frac{\epsilon^{\sf T}}{k_{\rm B} T}   \\
 & \quad \times \left(
\exp \frac{\mu^{\sf A}_{\rm S}}{k_{\rm B} T} - \exp \frac{\mu^{\sf B}_{\rm S}}{k_{\rm B} T}
\right) \,.
\end{split}
\ee
The central property of Eq \ref{eq:Flux_net_fundamental} is that the rate naturally vanishes when the chemical potentials $\mu_{\rm S}$ in the phases match at two-phase equilibrium. In other words, our analysis of the transition state occupancy and of the flux density naturally satisfies the principle of detailed balance.

\textit{Experimental signatures}
--
As our derivation is independent of the equations of state for the chemical potentials, the results apply to a range of materials, including specifically ideal as well as interacting solid solutions. Materials laws for the $\mu(x,T)$ can be readily inserted into Eq \ref{eq:Flux_net_fundamental} and the resulting expressions compared to experiment.

Specifically, electrochemistry provides experimental approaches to the kinetics. Assume that solute atoms in one of the phases -- without lack of generality, $\sf A$ -- equilibrate quasi instantaneously with ions of signed valency $z$ in an abutting electrolyte. For instance, $\sf A$ may be an absorption- or underpotential deposition layer. Then \cite{Kortuem1972}
\be
\label{eq:mu_vs_E}
\mu^{\sf A}_{\rm S} = - z q_0 (E-E_0)
\ee
($q_0$ - elementary charge, $E$ - electrode potential, $E_0$ - suitable reference potential). The associated variation in $\mathcal{N}_{\rm S}$ -- and, hence, the flux $J$ -- can be traced by monitoring the transferred charge, $\delta Q =   - z q_0  \delta \mathcal{N}_{\rm S}$. At stationary $E$, the system evolves towards $\mu^{\sf B}_{\rm S} = \mu^{\sf A}_{\rm S}$, which is the chemical potential at equilibrium, here controlled by $E$.

When exploring the dependency of the kinetics on $\mu^{\sf A}_{\rm S}$, a $\mu_{\rm S}$-dependence of $\epsilon^{\sf T}$ must be admitted, since the Brønsted-Evans-Polanyi principle \cite{VanSanten2009} suggests that varying the energies of the stable states will entail a variation of the transition state energy. Without committing to a specific functional form, we here consider quite generally that
\be
\label{eq:transfer_coefficient_generic}
\epsilon^{\sf T}
=
\epsilon^{\sf T}(\mu^{\sf A}_{\rm S},\mu^{\sf B}_{\rm S}) \,.
\ee
Note that the function $\epsilon^{\sf T}(\mu^{\sf A}_{\rm S},\mu^{\sf B}_{\rm S})$ cannot be expected linear. While generic entropy-of-mixing trends let the chemical potentials diverge in the limits of high dilution and approach to saturation, a divergence of $\epsilon^{\sf T}$ -- which would then be suggested if Eq \ref{eq:transfer_coefficient_generic} was a linear function -- is counterintuitive.

\begin{figure*} [!t]
	\centering
	\includegraphics[width=0.95 \textwidth]{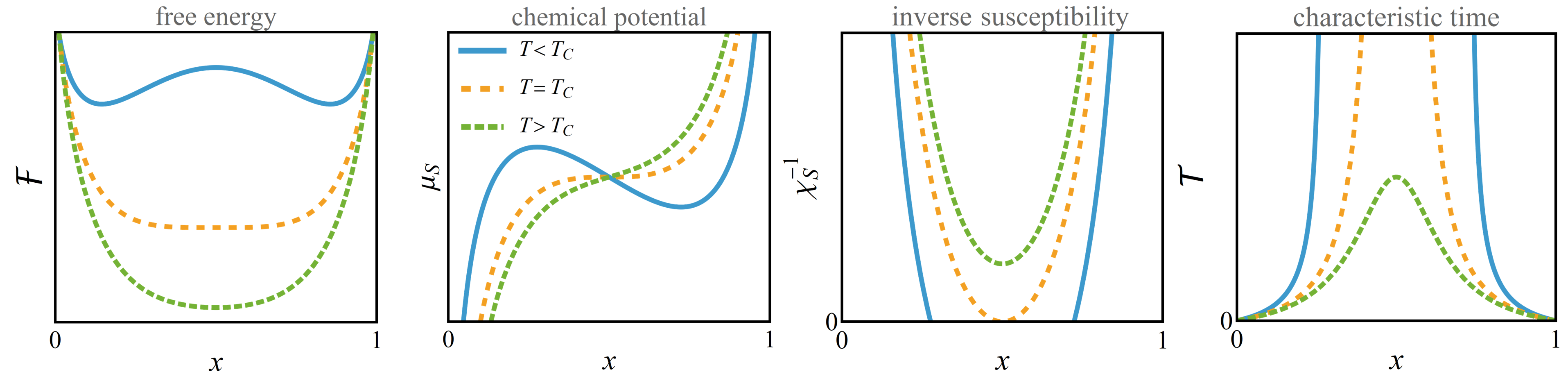}
	\caption{Schematic diagrams showing the solute fraction ($x$) dependence of free energy, $\mathcal F$, solute chemical potential, $\mu_{\rm S}$, inverse solute susceptibility, $\chi^{-1}_{\rm S}$, and characteristic charging time, $\mathcal T$, in response to an infinitesimal increment in $\mu_{\rm S}$. Schematic represents a solid solution with a miscibility gap (as apparent from the double-well free energy function) at temperatures $T$ below a critical temperature $T_{\rm C}$. Graphs refer to $T$ below, at and above $T_{\rm C}$, as indicated in legend. Below $T_{\rm C}$, extrema in $\mu_{\rm S}$ locate the spinodals, where $\chi^{-1}_{\rm S} \rightarrow 0$ and, in the consequence, $\mathcal T$ diverges. Solid solution is instable between the spinodals.}
	\label{fig:susceptibility}
\end{figure*}

In electrochemistry, an experimental approach to the exchange current density measures the response of the current density to small variations in $E$ around the equilibrium value \cite{Gerischer1957,Lasia2019}. The analogue in the present analysis is a flux resistance parameter in the form of the response, $J^{\rm eq}_{\mu} = \dd J^{\sf B} / \dd \mu^{\sf A}_{\rm S} \rvert_{T}$, of $J^{\sf B}$ to small variations in $\mu^{\sf A}_{\rm S}$ around the equilibrium value $\mu^{\rm eq}_{\rm S}$. $J^{\rm eq}_{\mu}$ characterizes the kinetics and it sets a boundary condition for bulk diffusion fields \cite{WuerschumPdH2024}.

Starting out with Eq \ref{eq:Flux_net_fundamental} and accounting for Eq \ref{eq:transfer_coefficient_generic} for $\epsilon^{\sf T}$, we obtain$J^{\rm eq}_{\mu}$ as the $\mu^{\sf A}_{\rm S}$-derivative of $J^{\sf B}$, evaluated at $\mu^{\sf A}_{\rm S} = \mu^{\sf B}_{\rm S} = \mu^{\rm eq}_{\rm S}$. The $\mu_{\rm S}$-derivative of $\epsilon^{\sf T}$ drops out and one obtains
\be
\label{eq:J_mu}
J^{\rm eq}_{\mu}
=
\frac{\nu_0 \rho_{\sf T} }{k_{\rm B} T}
(1 - x^{\sf A}) (1 - x^{\sf B})
  \exp \frac{\mu^{\rm eq}_{\rm S} - \epsilon^{\sf T}}{k_{\rm B} T} \,.
\ee
Except for the factor $1 / k_{\rm B} T$, this is readily seen to agree with the direction-resolved flux density, Eq \ref{eq:Flux_alpha_beta},evaluated at equilibrium--which is the exchange flux density.

Experiments may also investigate the time constant for relaxation into a new equilibrium state when, after equilibration at an initial value of $\mu^{\sf A}_{\rm S}$, the chemical potential of $\sf A$ is stepped by a small increment to a new stationary value \cite{BapariWeissmüller2025}.
Here, the present rate law yields (see Appendix)
\begin{gather}
\label{eq:J_equilibration}
J^{\sf B}
=
J^{\sf B}_{t = 0} \exp - \frac{t}{\mathcal T}
\\
\label{eq:T_equilibration}
{\mathcal T}^{-1}
=
\frac{A_{\rm V}}{\rho_0} \, \frac{J^{\rm eq}_{\mu}}{\chi^{\sf B}_{\rm S}} \,,
\end{gather}
where ${\mathcal T}$ denots the characteristic time for equilibration and $\chi^{\sf B}_{\rm S}$ represents a solute susceptibility \cite{LarcheCahnLinear1973} in $\sf B$,
\be
\label{eq:chi_definition}
\chi^{\sf B}_{\rm S}
=
\frac{\dd x^{\sf B}}{\dd \mu^{\sf B}_{\rm S}} \bigg \rvert_{T} \,.
\ee
The material-dependent function $\chi^{\sf B}_{\rm S}$ can vary strongly with $x$ (or $\mu^{\sf B}_{\rm S}$) and $T$.

Consider a solution with a miscibility gap. In Eq \ref{eq:T_equilibration}, the chemical potentials vary continuously in the vicinity of the miscibility gap critical point, hence $J^{\rm eq}_{\mu}$ is also continuous. Yet $\chi_{\rm S}^{\sf B}$ diverges at the critical point, as schematically illustrated in Fig \ref{fig:susceptibility}. Then, according to Eq \ref{eq:T_equilibration}, ${\mathcal T}$ must also diverge (${\mathcal T}^{-1}$ vanishes) there. More generally, the condition $\chi_{\rm S}^{\sf B} \rightarrow \infty$ defines the spinodals in $\sf B$, hence ${\mathcal T}$ also diverges at the spinodals. As the experiments in \cite{BapariWeissmüller2025} explore the charging time constant near the critical point but at a somewhat more elevated temperature, the strongly increased ${\mathcal T}$ near criticality in \cite{BapariWeissmüller2025} agrees well with the present theory.

\textit{Discussion}
--
As the subject of this work, the kinetics of particle transfer across crystal interfaces is of interest to phase transformations in solids and to the particle insertion into interstitial solutions and intercalation compounds. In particular, we have advertised the connection to electrochemical studies of the charging/discharging kinetics of metal hydrides.

Even though we explore the kinetics of charge-neutral particles as the rate-limiting process, electrochemistry affords monitoring that process when it is coupled to a fast (not rate-limiting) redox reaction. It is then of interest to compare our central result, the rate law Eq \ref{eq:Flux_net_fundamental}, to the BV equation \cite{Dickinson2020,Fletcher2023} as a frequently used rate law in electrochemistry. In view of Eq \ref{eq:mu_vs_E}, that equation may be expressed -- in terms of the particle flux density $J$ and the chemical potentials -- as
\be
\label{eq:Butler_Volmer}
J
=
J_0
\left(
\exp \frac {(1-\alpha) \Delta \mu_{\rm S}}{k_{\rm B} T}
-\exp \frac {- \alpha \Delta \mu_{\rm S}}{k_{\rm B} T}
\right)\, ,
\ee
with $J_0$ a constant exchange flux density parameter, $\alpha$ the transfer coefficient, and $\Delta \mu_{\rm S} = \mu^{\sf A}_{\rm S} - \mu^{\sf B}_{\rm S}$.
Appendix B compares implications of the present approach to those of the BV equation. Most notably, the present approach naturally reproduces an asymmetry -- revealed by experiment \cite{BapariWeissmüller2025} and without equivalent in the BV approach -- between the $\mathcal T$ on both sides of the critical point.
The rate law of the present work is distinguished by the explicit variation of its exchange flux density with the composition or chemical potential at equilibrium. That distinction rests on the $\mu$ in the phases entering separately and explicitly in Eq \ref{eq:Flux_net_fundamental}, whereas Eq \ref{eq:Butler_Volmer} involves exclusively their difference.

At the heart of the above distinction is the difference in premises. Equation \ref{eq:Butler_Volmer} describes the exchange between sites of a priori asserted occupancy or occupancy probability. Activating the transition state then involves the energy change between ground- and transition state. By contrast, Eq \ref{eq:Flux_net_fundamental} admits the sites in fast particle exchange with neighborhoods in their respective phases. Transitions then do work against the chemical potentials in the phases. We analyze that scenario by treating the sites as coupled to ergodic canonical systems. We expect that the analysis remains valid if the transfer sites equilibrate only with neighborhoods that present statistically representative subvolumes of the respective phases.

\textbf{Acknowledgment}. Many thanks to Yuri Mishin, Gunther Wittstock and Roland Würschum for critical comments on the manuscript and to Seoyun Sohn and Jürgen Markmann for inspiring discussions.



\appendix

\begin{center}
	
\section*{Endmatter}

\end{center}

\section{Time constant}

Here, we derive Eq \ref{eq:T_equilibration} for $\mathcal{T}$, the time constant for relaxation into a new equilibrium state after an incremental jump in the chemical potential of $\sf A$.

In keeping with our constitutive assumptions, we consider fast diffusion and, hence, homogeneous solute fraction $x^{\sf B}$ throughout the bulk of ${\sf B}$ at all times, $t$. Then,
\be
\label{eq:SI_dxdt}
\frac{\dd x^{\sf B}}{\dd t} = J^{\sf B} \frac{A_{\rm V}}{\rho_0}
\ee
with $A_{\rm V}$ the volume-specific surface area and $\rho_0$ the bulk density of sites for interstitials. Furthermore, we have
\be
\label{eq:SI_djdt}
\frac{\dd J^{\sf B}}{\dd t}
=
 \frac{\dd x^{\sf B}}{\dd t} \,
 \frac{\dd J^{\sf B}}{\dd x^{\sf B}} \bigg \rvert_{\mu^{\sf A}_{\rm S} = \rm{const}, \, \mu^{\sf B}_{\rm S} = \mu^{\sf A}_{\rm S}} \,.
 \ee
Analogously to the derivation of Eq \ref{eq:J_mu}, the derivative $J_x = \dd J^{\sf B}/\dd x^{\sf B}$ in Eq \ref{eq:SI_djdt} can again be based on Eq \ref{eq:W_total_closed_explicit} with Eq \ref{eq:transfer_coefficient_generic} for $\epsilon^{\sf T}$. The $\mu_{\rm S}$-derivative of $\epsilon^{\sf T}$ drops out, and one obtains
\be
\begin{split}
\label{eq:SI_djdx}
J_x
&=
\frac{\dd J^{\sf B}}{\dd x^{\sf B}} \bigg \rvert_{\mu^{\sf A}_{\rm S} = \rm{const}, \, \mu^{\sf B}_{\rm S} = \mu^{\sf A}_{\rm S}} \\
&=
\frac{\dd J^{\sf B}}{\dd \mu^{\sf B}_{\rm S}} \bigg \rvert_{\mu^{\sf A}_{\rm S} = \rm{const}, \, \mu^{\sf B}_{\rm S} = \mu^{\sf A}_{\rm S}}  \,
\frac{\dd \mu^{\sf B}_{\rm S}}{\dd x^{\sf B}}\\
&=
-\frac{\dd J^{\sf B}}{\dd \mu^{\sf A}_{\rm S}}
\bigg \rvert_{\mu^{\sf B}_{\rm S} = \rm{const}, \,
\mu^{\sf A}_{\rm S} = \mu^{\sf B}_{\rm S}}   \,
\frac{\dd \mu^{\sf B}_{\rm S}}{\dd x^{\sf B}} \\
&=
-\frac{
J^{\rm eq}_{\mu}}{\chi^{\sf B}_{\rm S}} \,,
\end{split}
\ee
where $\chi^{\sf B}_{\rm S}$ is the solute susceptibility of Eq \ref{eq:chi_definition} and the flux resistance parameter $J^{\rm eq}_{\mu}$ is defined (see main text) as
\be
\label{eq:SI_J_mu}
J^{\rm eq}_{\mu}
=
\frac{\dd J^{\sf B}}{\dd \mu^{\sf A}_{\rm S}} \bigg \rvert_{\mu^{\sf B}_{\rm S} = \rm{const}, \, \mu^{\sf A}_{\rm S} = \mu^{\sf B}_{\rm S}}  \,.
\ee
Note that the third line of Eq \ref{eq:SI_djdx} replaces the $\dd J^{\sf B}/\dd \mu^{\sf B}_{\rm S}$ derivative with $J^{\rm eq}_{\mu}$ of Eq \ref{eq:SI_J_mu}. In view of Eq 15, first changing the variable of differentiation from $\mu^{\sf B}_{\rm S}$ to $\mu^{\sf A}_{\rm S}$ and then accounting for $\mu^{\sf B}_{\rm S} = \mu^{\sf A}_{\rm S}$ at equilibrium leaves the magnitude of the derivative invariant but inverts its sign.

Since our attention is here on small jumps, $\delta \mu_{\rm S}$, in $\mu_{\rm  S}$ around its current value, all quantities in Eqs \ref{eq:SI_dxdt} - \ref{eq:SI_djdx} -- and, hence, $J_x$ -- may be taken as constants when studying the response to the jump.

Combining the above expressions, one obtains the differential equation
\be
\label{eq:SI_J_vs_t_DEQ}
\frac{1}{J^{\sf B}}
\frac{\dd J^{\sf B}}{\dd t}
=
\frac{A_{\rm V}}{\rho_0} J_x \,,
\ee
which is readily integrated, yielding
\begin{gather}
\nonumber
J^{\sf B}
=
J^{\sf B}_{t = 0}  \exp - \frac{t}{\mathcal T}
\\
\nonumber
{\mathcal T}^{-1}
=
- \frac{A_{\rm V}}{\rho_0} J_x
=
+\frac{A_{\rm V}}{\rho_0} \, \frac{J^{\rm eq}_{\mu}}{\chi^{\sf B}_{\rm S}} \,,
\end{gather}
which are Eqs \ref{eq:J_equilibration} and \ref{eq:T_equilibration} of the main text. ${\mathcal T}$ is the characteristic time for equilibration and $J^{\sf B}_0$ a constant (independent of $t$) prefactor. With the definition of $J^{\rm eq}_{\mu}$ in Eq \ref{eq:SI_J_mu}, the current at $t = 0$ is $J^{\sf B}_{t = 0}$
\be
J^{\sf B}_{t = 0}
=
J^{\rm eq}_{\mu} \, \delta \mu_{\rm S} \,.
\ee

\begin{figure*} [!t]
	\centering
	\includegraphics[width=0.95 \textwidth]{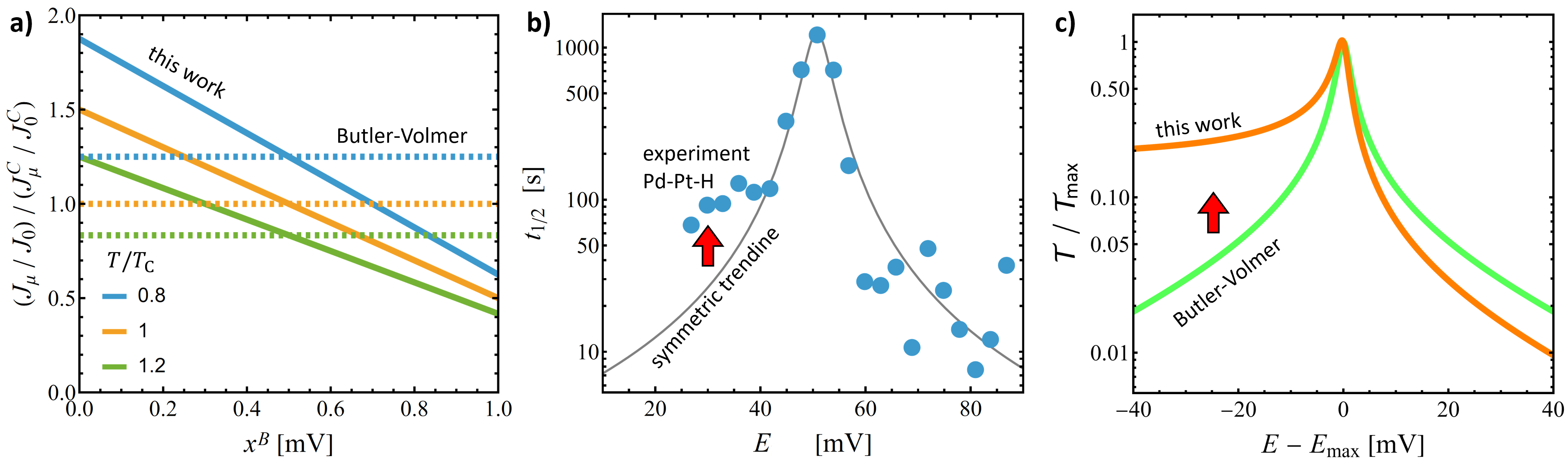}
	\caption{Implications of Butler-Volmer (BV) law and present approach, for an interacting regular solution in contact with a nearly saturated solution. (a), current-potential response, $J_{\mu}$, versus solute fraction, $x$. All graphs are  normalized to their respective prefactor ($J_0$ or $\tilde J_0$) and to their value ($J^{\rm C}_{\mu}/J^{\rm C}_0$ or $J^{\rm C}_{\mu}/\tilde J^{\rm C}_0$) at the regular solution critical point. Solid lines, present approach; dashed lines, BV law. Color code and legend: graphs for three different temperatures $T$, specified relative to the critical temperature $T_{\rm C}$.  (b), experimental \cite[Fig 12]{BapariWeissmüller2025} halftime, $t_{1/2}$, for H equilibration in nanoporous Pd-Pt-H after incremental jumps in the electrode potential, $E$, at $T$ just above $T_{\rm C}$. Blue dots: data; gray line: trendline symmetric about the maximum. (c), predictions of present (orange line) and BV (green line; symmetric around maximum) approach for halftime ${\mathcal T}$, here at $T = 1.1 \, T_{\rm C}$. Axes are referred to values at maximum; note identical axes range span as in (b). Red arrows in (b) and (c) mark asymmetry, larger halftime on the low-$E$ (high-$x$) side of the maximum.}
	\label{fig:comparison}
\end{figure*}

\section{Toy model and comparison to experiment}

\textit{Toy model.}
Here, we explore a toy model -- selected to incorporate the most important physics in the simplest possible way and with a minimal set of materials parameters -- for the materials laws that illuminates the distinction between the implications of the present approach and those of the BV law, including a qualitative comparison to experimental data.

For the materials equation of state we consider an interstitial regular solution with a parabolic (in $x$) attractive solute-solute interaction energy, qualitatively representative of solutions with a miscibility gap, as in many interstitial systems. We have in mind a core-shell particle with the shell a nearly saturated solution in the parameter range of interest; such scenarios are found in metal hydrides and lithium compounds \cite{Jerkiewicz2010,CogswellBazant2013,LiWeissmüller2025Coherency}. The molar free energy of the regular solution is
\be
\label{eq:SI_g_reg_sol}
f
=
\mu_0 \, x
+ \omega \, x \, (1-x)
+ k_{\rm B} T \left[ x \ln x + (1-x) \ln (1-x) \right]
 \,,
\ee
with $\omega$ (here $>0$) and $\mu_0$ solute-solute and solute-matrix interaction parameters, respectively. Equation \ref{eq:SI_g_reg_sol} implies the chemical potential
\begin{gather}
\label{eq:SI_mu_S_reg_sol_EOS}
\mu_{\rm S} = \mu_0 + \omega \, (1 - 2 x)
+ k_{\rm B} T \ln \frac x {1-x} \,,
\end{gather}
and the solute susceptibility
\be
\chi_{\rm S}
=
\frac{x (1-x)}{k_{\rm B} T -  2 \omega x (1-x)} \,,
\ee
which diverges at the spinodals and at the critical point, $x = \tfrac 1 2$ and $T = T_{\rm C} = \omega / (2R)$.

For the shell, in the approach to saturation ($x \rightarrow 1$), $\mu$ varies asymptotically as
\begin{gather}
\label{eq:SI_mu_S_sat}
\mu^{\rm sat}_{\rm S} =
\mu_0 - \omega
- k_{\rm B} T \ln (1-x) \,;
\end{gather}
thus,
\begin{gather}
\label{eq:SI_x_S_sat}
x^{\rm sat}
=
1 - \exp -\frac{\mu - \mu_0 + \omega}{k_{\rm B} T} \,.
\end{gather}

\textit{Current-potential derivatives.}
We now insert the above results into Eq \ref{eq:J_mu} for the current-potential derivative at equilibrium. Taking specifically Eqs \ref{eq:SI_mu_S_reg_sol_EOS} for the core (bulk, $\sf B$, with $\mu_0^{\sf B} = 0$ as a reference convention) and \ref{eq:SI_mu_S_sat}, \ref{eq:SI_x_S_sat} for the shell ($\sf S$), and accounting for the equilibrium condition $\mu^{\sf S, \rm eq}_{\rm S} = \mu^{\sf B, \rm eq}_{\rm S}$ yields
\be
\label{eq:SI_J_mu_toy}
J^{\rm eq}_{\mu}
=
\frac{\tilde J_0}{k_{\rm B} T}
\left( 1 - x^{\sf B} \right)\,
\ee
with the temperature-dependent prefactor
\be
\label{eq:SI_J_mu_toy_prefactor}
\tilde J_0
=
\nu_0 \rho_{\sf T} \,
  \exp - \frac{\epsilon^{\sf T} - \mu_0^{\sf S} + \omega^{\sf S}  }{k_{\rm B} T} \,.
\ee
In the limiting case where the transition state energy is independent of the composition, this reduces to a linear composition dependence of the exchange current density.

$J^{\rm eq}_{\mu}$ is also readily evaluated for the BV equation, Eq \ref{eq:Butler_Volmer}; the result is
\be
\label{eq:SI_J_mu_BV}
J^{\rm eq, BV}_{\mu}
=
\frac{J_0 }{k_{\rm B} T} \,.
\ee
Figure \ref{fig:comparison}(a) compares the two above expressions for $J_{\mu}$, as a function of the core solute fraction $x$ and for different values of the reduced temperature $T / T_{\rm C}$. The figure assumes a constant, composition-independent transition state energy. Contrary to the constant exchange current of the BV equation, the present approach reveals an explicit composition-dependence of the current-potential derivative. The variation is moderate near the center of the composition interval, and it is strong in the approach to saturation.

Recall that this result is specific for the toy model, regular solution core and nearly saturated shell. More general behavior may be expected with different core-shell combinations and more realistic equations of state for the free energies.

\textit{Characteristic time for equilibration. }
Equations \ref{eq:SI_J_mu_toy} and \ref{eq:SI_J_mu_BV} for the $J^{\rm eq}_{\mu}$ can be combined with Eq \ref{eq:T_equilibration} in order to compute the characteristic time for equilibration. The results are
\be
\label{eq:SI_T_regsol}
\mathcal{T}
=
\frac{\rho_0}{A_{\rm V}} \,
\frac{k_{\rm B} T}{\tilde J_0} \,
\frac{x^{\sf B}} {k_{\rm B} T - 2 x^{\sf B} (1-x^{\sf B}) \omega^{\sf B}}
\ee
for the present approach and
\be
\label{eq:SI_T_BV}
\mathcal{T}^{\rm BV}
=
\frac{\rho_0}{A_{\rm V}} \,
\frac{k_{\rm B} T}{J_0} \,\,
\frac{x^{\sf B} (1-x^{\sf B})} {k_{\rm B} T - 2 x^{\sf B} (1-x^{\sf B}) \omega^{\sf B}} \,\,
\ee
for BV.

Figure \ref{fig:comparison} (b) shows experimental results for $\mathcal T$ of the H equilibration in nanoporous Pd-Pt-H at a temperature slightly above $T_{\rm C}$ \cite[Fig 12]{BapariWeissmüller2025}. Noteworthy observations are a peak near the critical composition and a pronounced asymmetry of the $\mathcal T$ on the two sides of that peak.

Figure \ref{fig:comparison} (c) displays the predictions of Eqs \ref{eq:SI_T_regsol} and \ref{eq:SI_T_BV}. For qualitative comparability with the experimental results, the abscissa and ordinate span the identical intervals as in panel (b). For contact to the experiment, the acting and critical temperatures in Eq \ref{eq:SI_T_regsol} were set to $T = 1.1 T_{\rm C}$ and $T_{\rm C} = \rm 270 K$. The remaining parameters drop out in the display of normalized quantities. The solute fraction $x$ was converted to chemical potential and then to electrode potential $E$ by means of Eqs \ref{eq:SI_mu_S_reg_sol_EOS} and \ref{eq:mu_vs_E}. Comparing Figs \ref{fig:comparison} (b) and (c) and accounting for the qualitative nature of the comparison, one perceives a good overall agreement.

In Fig \ref{fig:comparison} (c), the variation of the $\mathcal T$ with $E$, and specifically the peak near the critical composition, reflects primarily the solute susceptibility. On top of that variation, one notes the distinction that the BV-based law is symmetric around the peak, whereas the present approach predicts a strong asymmetry (up to a full order of magnitude within the parameter set of the figure), larger equilibration time at smaller $E$, hence in more concentrated solutions.

Comparing the two approaches, BV and the present one, to experiment and theory, Fig \ref{fig:comparison} (b) and (c), one finds that the present approach is qualitatively closer to the experiment inasmuch as it correctly and naturally embodies the pronounced asymmetry in the experimental time constants for compositional equilibration.

\end{document}